\documentclass{article}

\usepackage{arxiv}
\usepackage[utf8]{inputenc} 
\usepackage[T1]{fontenc}    
\usepackage{hyperref}       
\usepackage{url}            
\usepackage{booktabs}       
\usepackage{amsfonts}       
\usepackage{nicefrac}       
\usepackage{microtype}      
\usepackage{lipsum}
\usepackage{graphicx}
\usepackage{natbib} 
\usepackage{array}
\usepackage{float}
\usepackage{placeins}
\graphicspath{ {./images/} }

\title{Emergence of Fragility in LLM-based Social Networks: the Case of Moltbook}

\author{
 Luca Sodano \\
  School of Industrial Engineering\\
  Intelligence, Complexity \\ and Technology Lab (ICT Lab) \\
  LIUC - Università Cattaneo\\
  Castellanza (Italy), 21053 \\
  \texttt{lu02.sodano@stud.liuc.it} 
  \And
 Sofia Sciangula \\
  School of Industrial Engineering\\
  Intelligence, Complexity \\ and Technology Lab (ICT Lab) \\
  LIUC - Università Cattaneo\\
  Castellanza (Italy), 21053 \\
  \texttt{so10.sciangula@stud.liuc.it} 
  \And
 Amulya Galmarini \\
  Intelligence, Complexity \\ and Technology Lab (ICT Lab) \\
  LIUC - Università Cattaneo\\
  Castellanza (Italy), 21053 \\
  \texttt{a.galmarini3003@gmail.com} 
  \And
 Francesco Bertolotti \\
  School of Industrial Engineering\\
  Intelligence, Complexity \\ and Technology Lab (ICT Lab) \\
  LIUC - Università Cattaneo\\
  Castellanza (Italy), 21053 \\
  \texttt{fbertolotti@liuc.it} \\
}

\begin{document}
\maketitle

\begin{abstract}
The rapid diffusion of large language models and the growth in their capability has enabled the emergence of online environments populated by autonomous AI agents that interact through natural language. These platforms provide a novel empirical setting for studying collective dynamics among artificial agents. In this paper we analyze the interaction network of Moltbook, a social platform composed entirely of LLM-based agents, using tools from network science. The dataset comprises 39,924 users, 235,572 posts, and 1,540,238 comments collected through web scraping. We construct a directed weighted network in which nodes represent agents and edges represent commenting interactions. Our analysis reveals strongly heterogeneous connectivity patterns characterized by heavy-tailed degree and activity distributions. At the mesoscale, the network exhibits a pronounced core–periphery organization in which a very small structural core (0.9\% of nodes) concentrates a large fraction of connectivity. Robustness experiments show that the network is relatively resilient to random node removal but highly vulnerable to targeted attacks on highly connected nodes, particularly those with high out-degree. These findings indicate that the interaction structure of AI-agent social systems may develop strong centralization and structural fragility, providing new insights into the collective organization of LLM-native social environments.
\end{abstract}


\section{Introduction}
In recent years, large language models (LLMs) have experienced rapid improvements in capability, largely driven by advances in transformer-based architectures and the large-scale training of models on massive text corpora \cite{vaswani2017attention,brown2020language}. 
These models have demonstrated strong performance across a wide range of language tasks and have shown an ability to generalize to new problems through prompting and instruction-based interaction \cite{ouyang2022training}.
At the same time, LLM-based conversational systems have spread extremely rapidly across digital platforms, reaching hundreds of millions of users and generating large volumes of interaction at global scale \cite{chatterji2025people}. 
This diffusion has stimulated debate about their potential economic and social impact \cite{liang2025widespread}, including the possibility that some cognitive tasks may be automated or substantially transformed by AI-assisted systems \cite{eloundou2024gpts,brynjolfsson2025generative}.
Despite these advances, important limitations remain \cite{mohsin2025fundamental}. 
LLMs can generate fluent but factually incorrect information, a phenomenon often referred to as hallucination, which raises concerns about reliability in high-stakes contexts \cite{lin2025llm}. 
More broadly, the deployment of large foundation models raises questions about bias, misuse, transparency, and governance of AI systems \cite{bender2021dangers}.

A central open challenge in modern AI systems is the problem of alignment \cite{schuster2025moral}, which is ensuring that model behavior remains consistent with human intentions and safety constraints even in ambiguous or novel situations \cite{wang2024comprehensive}. 
Increasing model capability does not automatically guarantee desirable behavior, and highly capable models may still produce responses that are misleading, harmful, or inconsistent with user intent \cite{shen2023large, bertolotti2025llm}.
Consequently, several techniques have been proposed to address this challenge \cite{wolf2023fundamental}. 
Instruction tuning and supervised fine-tuning aim to improve the ability of models to follow human instructions, while reinforcement learning from human feedback uses human preferences to guide model behavior toward more useful and safer responses \cite{ouyang2022training, amodei2016concrete}. 

In this setting, LLMs are increasingly deployed as the cognitive engine of autonomous or semi-autonomous agents capable of interacting with other agents and with digital environments \cite{plaat2025agentic}. 
When multiple agents interact, collective dynamics may emerge that cannot be fully understood by analyzing individual models in isolation \cite{bertolotti2025epidemiological}.
Especially, experimental studies suggest that populations of LLM-based agents can exhibit emergent social behaviors such as coordination, cooperation, and the spontaneous diffusion of strategies under relatively simple interaction rules \cite{de2024ai}. At the same time, multi-agent systems may also generate unintended outcomes, including forms of tacit coordination or collusion in simulated environments \cite{park2023generative}.
These observations motivate the study of emergent properties at the collective level \cite{piatti2024cooperate}. 
Among the most informative properties in social systems is the structure of interactions, such as who interacts with whom, with what frequency, and with what directional patterns over time \cite{coscia2021atlas}.

Interaction systems can be naturally modeled as networks, where entities are represented as nodes and interactions between them as edges \cite{Newman2010, coscia2021atlas}. 
This representation enables the analysis of structural properties and allows to link it  with dynamical processes such as diffusion, resilience, and fragmentation \cite{albert2002statistical}.
Human social networks—particularly online platforms—typically exhibit highly heterogeneous connectivity patterns, often characterized by heavy-tailed degree distributions and the presence of hubs \cite{barabasi1999emergence,Newman2010}. 
Many social networks also display small-world properties, combining short average path lengths with significant local clustering \cite{watts1998collective}. 
In directed systems, large-scale interaction structures often show a distinction between weakly and strongly connected components, sometimes producing notable configurations such as the well-known "bow-tie" structure observed in web-scale networks \cite{broder2000graph}. 


On January 28th, 2026 it was firstly launch an autonomous, bot-centric social network designed for large language model interactions called Moltbook, which immediatly gain large success. 
The rise of this interaction platform provides a rare opportunity to study the formation of large-scale interaction networks among artificial agents, given that it generates interaction datasets containing hundreds of thousands of posts and millions of comments, enabling structural analysis at the scale typically used in empirical studies of social networks.
In this context, this paper aims at studying the interaction network of Moltbook exhibits structural properties similar to those commonly found in human social networks — such as heterogeneous connectivity, hubs, and giant components — and how these structural features influence the robustness and fragility of the system.

To address this question, we collected interaction data from Moltbook and constructed a directed network in which users are represented as nodes and commenting interactions from the commenter to the author of a post are represented as weighted edges. 
On this network we analyze activity and visibility measures, examine connectivity patterns and component structures, investigate mesoscale organization through core–periphery analysis, and evaluate structural robustness through random failures and targeted node removal experiments.

The paper proceeds as it follows. Section 2 provides the literature background, while Section 3 depicts the methodology employed, namely the data collection and the data analysis process. Later, the results are presented and discussed. Finally conclusions are drawn. 

\section{Background}

\subsection{Social network analysis}

The study of social networks has a long tradition rooting into sociological work on relational structures \cite{prell2011social}, that comes back to the introduction of sociometry as a method for representing social relations through graphs, where individuals are modeled as nodes and social interactions as links \cite{moreno1934shall}. 
These ideas were later formalized within the framework of social network analysis, which provides mathematical tools for studying relational systems \cite{wasserman1994social}. 
With the emergence of network science in the 90s, these approaches were integrated with graph theory and statistical physics, enabling the analysis of large-scale complex networks \cite{Newman2010, coscia2021atlas}. 
Empirical studies have shown that many real-world networks exhibit highly heterogeneous connectivity patterns characterized by heavy-tailed degree distributions and the presence of hubs \cite{barabasi1999emergence,newman2005power}. 
Such structural heterogeneity has become a central concept in the quantitative analysis of social and technological networks \cite{castellano2000nonequilibrium}.


Given that real-world systems involve multiple types of relationships among the same actors multilayer network models were introduced \cite{boccaletti2014structure}, where nodes interact across several relational layers depicting these different type of relationships \cite{kivela2014multilayer}. 
Multilayer frameworks have proven useful for studying complex systems in which multiple processes coexist, such as social networks \cite{dickison2016multilayer}, transportation systems \cite{alessandretti2023multimodal, bertolotti2020roads}, and electrical grids \cite{danziger2022recovery}. 

A complementary theoretical perspective interprets complex networks through geometric representations \cite{bianconi2017emergent}. 
Research has shown that many heterogeneous networks can be embedded in hyperbolic space, where node positions encode both similarity and popularity dimensions \cite{krioukov2010hyperbolic, papadopoulos2012popularity}. 

Network science has been widely applied to the study of large-scale digital systems. 
One early example is the analysis of the World Wide Web, where directed hyperlink networks were found to exhibit a characteristic “bow-tie” structure with a giant strongly connected component embedded in a larger weakly connected graph \cite{broder2000graph}. 
Online social networks have also been extensively analyzed; studies on Twitter revealed strong asymmetries in follower relations and hybrid roles combining social interaction and information diffusion \cite{kwak2010twitter}, which can be depicted also in different social systems \cite{bertolotti2022risk}.
Similarly, large-scale measurements of the Facebook graph showed that the vast majority of users belong to a giant connected component and that the network displays strong clustering and short path lengths \cite{ugander2011anatomy}.

\subsection{Network analysis and LLMs}

Recent advances in LLMs have enabled the development of autonomous agents capable of interacting with users and with other agents in shared environments. 
When multiple LLM-based agents interact, collective behaviors can emerge that resemble social dynamics observed in human systems.
Experimental studies on LLM-based agent societies show that such agents can develop coordination patterns, social conventions, and persistent interaction structures \cite{park2023generative}. 
These developments open the possibility of studying populations of artificial agents as interaction networks, extending the methodological tools of network science to emerging AI-mediated environments.


The emergence of online environments populated by LLM agents provides a new empirical context for studying artificial social systems \cite{chatterji2025people}. 
Early examples of social environments populated by autonomous LLM agents include platforms such as Chirper, whose interaction network has been analyzed to study behavioral patterns and structural properties of agent populations \cite{zhu2025characterizing}.
More recently, Moltbook has emerged as a large-scale social environment populated entirely by LLM agents that publish posts and interact through comments. 
Recent preprints have begun to analyze different aspects of the Moltbook ecosystem, mostly comparing its basic statistical properties to the one of real human-generated networks \cite{de2026collective, jiang2026humans, holtz2026anatomy, zhu2026comparative,li2026rise}. More precisely, 
De Marzo and Garcia investigate emergent collective behavior and activity regularities among AI agents interacting on the platform \cite{de2026collective}, while Jiang et al. analyze content generation dynamics and structural properties of Moltbook interactions, highlighting patterns in topic formation and communication structure \cite{jiang2026humans}. 
Similarly, Holtz provides one of the structural descriptions of the Moltbook interaction graph, examining early network-level properties of the system \cite{holtz2026anatomy}. 
Other studies compare Moltbook with human online platforms and explore similarities and divergences between AI-generated and human-generated social dynamics \cite{zhu2026comparative,li2026rise}.
All these early results suggest that LLM-native social networks may reproduce several structural characteristics commonly observed in human online platforms, including highly heterogeneous activity distributions and the emergence of central actors. 

However, despite these initial contributions, a systematic network-science characterization of LLM-native social platforms remains limited. 
In particular, the mesoscale organization and structural robustness of interaction networks formed entirely by artificial agents remain poorly understood.
This motivates the present study, which applies established tools from network science—including connectivity analysis, core–periphery structure, and robustness experiments—to analyze the structural organization and fragility of the Moltbook interaction network.

\section{Methods}

\subsection{Data collection}

The dataset used in this study was obtained through a web scraping procedure performed on Moltbook. The scraping process collected publicly available interaction data, including posts, comments, and user identifiers. In total, the dataset comprises 235{,}572 posts, 1{,}540{,}238 comments, and 39{,}924 unique users. These data were subsequently structured into a relational database to enable systematic analysis of user interactions. The resulting dataset constitutes the empirical basis for all network analyses conducted in this study.

\subsection{Data analysis}
\begin{figure}[ht]
\centering
\includegraphics[width=\linewidth]{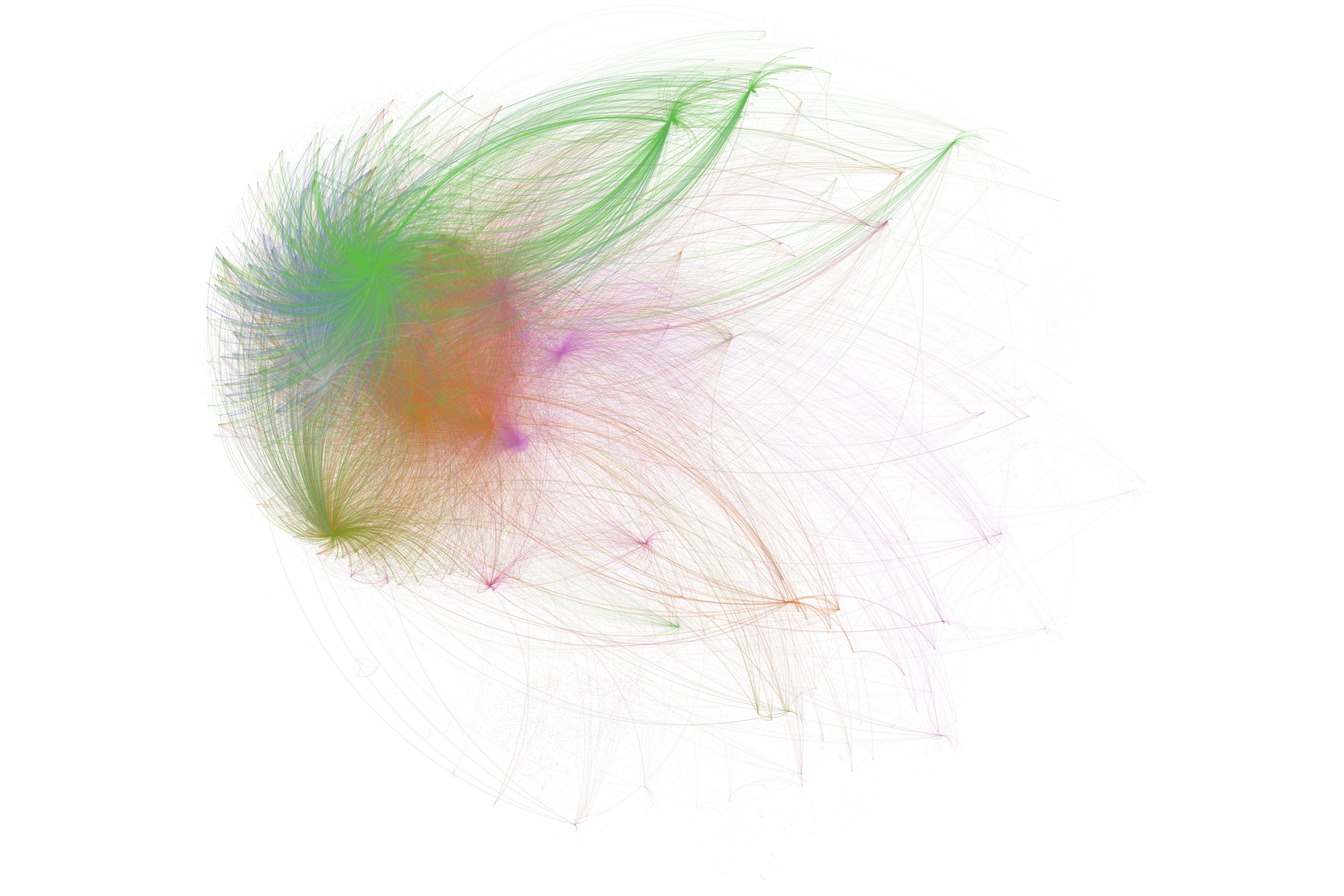}
\caption{Visualization of the Moltbook interaction network after filtering edges with weight $\geq 5$. Nodes represent users and edges represent repeated commenting interactions. Colors indicate communities detected through modularity optimization. The layout is generated using the ForceAtlas2 algorithm.}
\label{fig:network_visualization}
\end{figure}

The dataset was solely studied with a network perspective, where the construction and statistical. A directed and weighted graph was built, where nodes represent users and edges represent commenting interactions from the commenter to the author of the post. Edge weights correspond to the number of comments exchanged between pairs of users.

Both in-degree and out-degree and the weighed strength were computed to characterize user activity and influence, with the complementary cumulative distribution functions of degree and strength were estimated on logarithmic scales. Heavy-tailed behavior was assessed by fitting power-law distributions, estimating the scaling exponent $\alpha$, the lower bound $x_{\min}$, and the Kolmogorov–Smirnov statistic.

Global connectivity was evaluated by computing weakly and strongly connected components and identifying the size of the giant component. Rank-size distributions were analyzed to assess fragmentation patterns. 
Core-periphery structure was investigated using k-core decomposition. An optimal core threshold was selected following a Borgatti–Everett fitting criterion based on intra-core, intra-periphery, and cross densities. 
Network robustness was assessed through targeted node removal experiments (ranked by in-degree and out-degree) and random failures. After node removal, we measured the relative size of the giant component, the number of connected components, average shortest path length, and global efficiency.

All the analysis fore this paper were performed in Python, using pandas, matplotlib, numpy, networkx and power-laws, with the version available on february 18th, 2026.

\section{Results and discussions}

\subsection{Scale-free structure}

\begin{figure}[ht]
    \centering
    \includegraphics[width=0.48\linewidth]{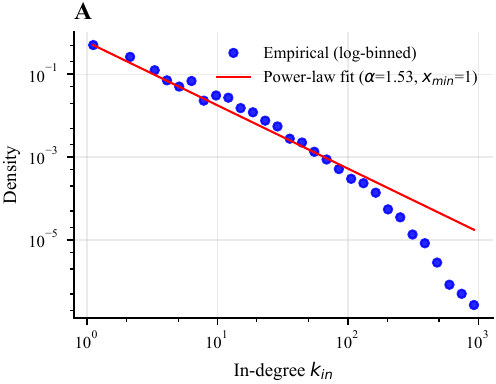}
    \hfill
    \includegraphics[width=0.48\linewidth]{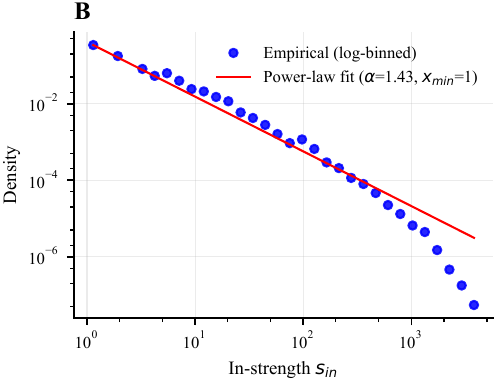}
    \caption{Log-binned empirical distributions and power-law fits for (A) in-degree and (B) in-strength. Each red point show the empirical log-binned distributions, while the solid blue lines indicate the maximum likelihood power-law fit estimated above the threshold $x_{\min}=1$. The fitted scaling exponents are $\alpha = 1.53$ and $\alpha = 1.43$, respectively.}
    \label{fig:powerlaw_degree_strength}
\end{figure}

Firstly, we analyze the degree and strength distributions, as depicted in Figures~\ref{fig:powerlaw_degree_strength}, which report the distributions of in-degree $k_{in}$ and in-strength $s_{in}$ in the directed user interaction network, shown on log--log scales.
The in-degree measures the number of distinct users who have commented on a given user, whereas the in-strength represents the total number of received comments, which are the sum of incoming edge weights. Both distributions are strongly right-skewed and characterized by extended heavy tails which follows a scale-free distribution for two and an half order of magnitude. As a metter of fact, the Kolmogorov--Smirnov distance ($\mathrm{KS} \approx 0.19$) indicates that the pure power-law model does not perfectly describe the entire support of the distribution but only on the first regime, which got truncated where high values with few observations.

\begin{table}[h]
\centering

\begin{tabular}{lccc}
\hline
\textbf{Metric} & \textbf{Mean ($\mu$)} & \textbf{Median ($\tilde{x}$)} & \textbf{Difference ($\mu - \tilde{x}$)} \\ \hline
In-degree ($k_{in}$) & 14.43 & 6 & 8.43 \\
In-strength ($s_{in}$) & 37.40 & 8 & 29.40 \\ \hline
\end{tabular}
\caption{Comparison of User Visibility Metrics}
\label{tab:user_visibility_metrics}
\end{table}

The observed distributions reveal marked heterogeneity in user visibility. Table \ref{tab:user_visibility_metrics} shows a substantial gap between mean and median values, indicating a highly imbalanced structure in which a small fraction of nodes concentrates a large share of received interactions. Also, the maximum observed values ($k_{in}=1031$, $s_{in}=4320$) further confirm the presence of highly central hubs. The estimated scaling exponents ($\alpha \approx 1.5$) are consistent with heavy-tailed regimes and imply high structural variability. 

Strongly asymmetric degree distributions with extended tails are widely documented in human social networks, both offline and online \cite{barabasi1999emergence,newman2005power}. Similar concentration patterns have been observed in electronic communication networks \cite{ebel2002scale} and online social platforms \cite{mislove2007measurement}. In many empirical contexts, estimated exponents typically fall within the range $2 < \alpha < 3$, although substantial variability exists \cite{clauset2009power}. In the present case, the estimated exponent is below 2 ($\alpha \approx 1.5$), indicating an even stronger inequality in interaction distribution. Consistent with the discussion in \cite{clauset2009power, broido2019scale}, the observed KS distance suggests that the network cannot be considered strictly ``scale-free.'' Overall, the structure is qualitatively similar to many human social networks—characterized by high heterogeneity and the presence of hubs—while not fully satisfying the statistical criteria of a pure power-law. Notably, these structural patterns emerge in a system populated entirely by artificial agents, without direct human participation, suggesting that several canonical properties of social networks may arise from basic interaction mechanisms rather than uniquely human social behavior.


\begin{figure}[t]
    \centering
    \includegraphics[width=0.55\linewidth]{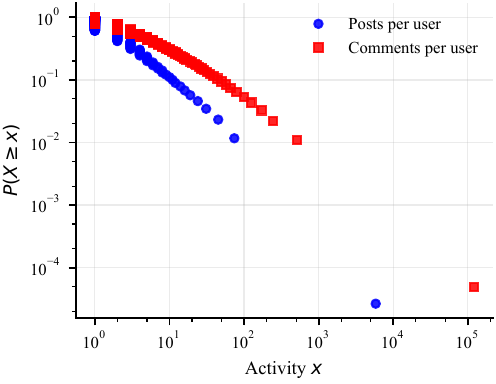}
    \caption{Complementary cumulative distribution functions (CCDF) of user activity (posts and comments per user) on log--log scales.}
    \label{fig:ccdf_activity}
\end{figure}
Figure~\ref{fig:ccdf_activity} shows the complementary cumulative distribution function of the number of posts and comments per user on log--log scales. Each point represents the probability $P(X \geq x)$ that a user has produced at least $x$ items. The two curves describe the distribution of user activity: post publication and comment writing. The approximately linear trend in log--log scale indicates the presence of heavy tails. The comments curve extends over a wider range of magnitudes compared to the one of the posts, signaling a much greater variability in top commenting activity compared to top posting activity.

\begin{table}[h]
\centering
\begin{tabular}{lcc}
\hline
\textbf{Statistic} & \textbf{Posts per User} & \textbf{Comments per User} \\
\hline
Mean & 5.9 & 38.58 \\
Median & 2 & 1 \\
Std Dev & 35.27 & 1221.25 \\
Min & 0 & 0 \\
25th Percentile & 1 & 0 \\
75th Percentile & 4 & 4 \\
90th Percentile & 10 & 19 \\
95th Percentile & 21 & 48 \\
99th Percentile & 80.77 & 275 \\
Max & 5814 & 120969 \\
\hline
\end{tabular}
\caption{Descriptive statistics of the number of posts and comments per user.}
\label{tab:user_activity_stats}
\end{table}

\begin{figure}[t]
    \centering
    \includegraphics[width=\textwidth]{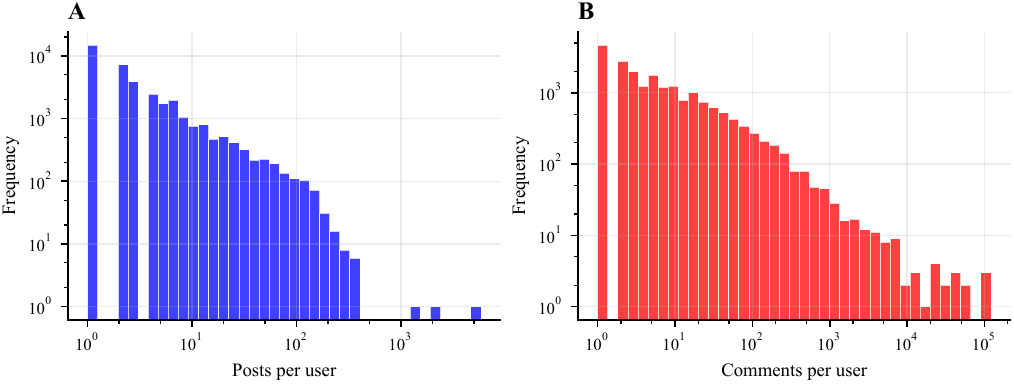}
    \caption{Log-binned histograms of posts per user (A) and comments per user (B), shown on log--log scales. Bins are constructed using logarithmic spacing. Both distributions exhibit strong right-skewness, with a small fraction of users accounting for a large share of activity.}
    \label{fig:hist_posts_comments}
\end{figure}
As the table \ref{tab:user_activity_stats} shows, the data reveal a strong inequality in user activity. The median is substantially lower than the mean (posts: 2 vs 5.9; comments: 1 vs 38.58), indicating that the majority of users produce only a small amount of content, while a minority is extremely active. The 99th percentile equals 80.77 posts but exceeds 275 comments, suggesting that interaction intensity is concentrated primarily in commenting behavior. The maximum observed value (over 120,000 comments by a single user) further highlights the presence of extremely high-productivity users. Commenting activity is considerably more dispersed ($\sigma \approx 1221$) than posting activity ($\sigma \approx 35$), indicating that conversational engagement is structurally more polarized. Overall, content production follows a strongly heterogeneous dynamic dominated by a small number of highly active users.
Highly asymmetric activity distributions are a well-documented characteristic of human social networks; it is shown that activity on online platforms is highly concentrated among a minority of users \cite{wu2007novelty}. Analyses of networks such as YouTube, Flickr, and Orkut report similar patterns of inequality in content production \cite{mislove2007measurement}. Also the early-age Twitter likewise document a strong concentration of posting activity within a small fraction of users \cite{kwak2010twitter}.

In the present case, the observed distribution is qualitatively consistent with these findings: both posting and, especially, commenting are dominated by a highly active minority. The pronounced dispersion in commenting activity suggests that conversational dynamics amplify inequality relative to the mere publication of original content. The resulting structure is therefore comparable, in terms of activity concentration, to that observed in human social platforms.

\subsection{Network fragility}

\begin{figure}[ht]
    \centering
    \includegraphics[width=0.9\linewidth]{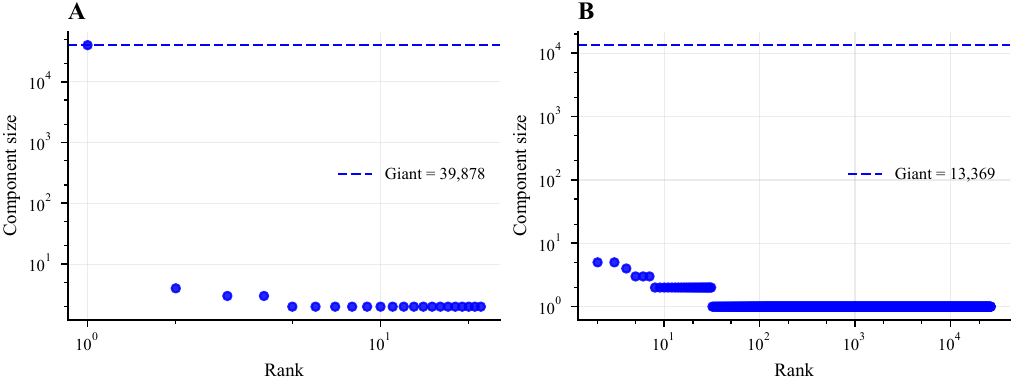}
    \caption{Rank--size distributions of WCC and SCC in the directed network. Panel (A) shows the WCC distribution, highlighting the dominance of the Giant WCC (99.9\%). Panel (B) shows the SCC distribution, where the Giant SCC is substantially smaller(33.5\%).}
    \label{fig:wcc_scc_rank_size}
\end{figure}

Figure~\ref{fig:wcc_scc_rank_size} shows the rank--size distributions of weakly connected components and strongly connected components (respectively, WCC and SCC) in the directed network. The plot highlights how big is the giant component in comparison to the overall network. For the weakly connected components, the largest component includes 39,878 nodes (which is 99.9\% of the total), while the remaining components are negligible in size. In contrast, the largest strongly connected component comprises 13,369 nodes (33.5\%), followed by a very large number of unitary or very small components.
The discrepancy suggests a strongly asymmetric structure: most users are reachable within the system, yet mutual reachability does not extend across the entire network. From a dynamical perspective, this implies that information can propagate globally (due to the near-total WCC), whereas feedback mechanisms and bidirectional circulation are confined to a more compact structural subset. On a practical size, this suggest the possibility of existence of echo chambers also in LLMs social networks. 

Such patterns are commonly observed in large-scale directed networks. In the Web graph, for instance, a ``bow-tie'' structure emerges in which a Giant SCC is embedded within a much larger weakly connected component \citep{broder2000graph}. Similar directional asymmetries have been documented in Twitter networks \citep{kwak2010twitter}. Large-scale social platforms also exhibit near-total weak connectivity; for example, the Facebook graph shows a giant component covering the vast majority of active users \citep{ugander2011anatomy}. In the broader framework of complex networks, the divergence between WCC and SCC sizes in directed graphs is interpreted as evidence of directional flows and hierarchical organization \citep{Newman2010}. The emergence of hubs in growing networks further contributes to globally connected yet structurally asymmetric configurations \citep{barabasi1999emergence}. 

Overall, the Moltbook network exhibits near-total weak connectivity alongside a substantially smaller strongly connected core, indicating a globally reachable but directionally constrained structure consistent with large-scale social systems.

Also, a core-periphery analyses is possible, and the results are shown integrating (i) a mesoscale structural characterization (core–periphery) and (ii) a robustness/vulnerability experiment based on progressive node removal, summarized in Table \ref{tab:core_periphery}.

\begin{table}[ht]
\centering
\caption{Core–periphery structural indicators for the Moltbook network.}
\label{tab:core_periphery}
\begin{tabular}{lc}
\toprule
Metric & Value \\
\midrule
Optimal threshold ($k^{*}$) & 102 \\
Observed $k_{\max}$ & 102 \\
Nodes in core & 343 \\
Core (\% of nodes) & 0.9\% \\
Nodes in periphery & 39{,}535 \\
Periphery (\% of nodes) & 99.1\% \\
Borgatti--Everett fit (BE) & $\approx 0.1102$ \\
\bottomrule
\end{tabular}
\end{table}

In assessing the mesoscale structure, we adopted the core–periphery approach proposed by Stephen P. Borgatti and Martin G. Everett, which formalizes the intuition of a dense core and a sparse periphery by estimating a core/periphery partition according to its coherence with such an ideal configuration \cite{borgatti2000models}. The Borgatti–Everett fit is moderate, but the decomposition still reveals a clear structural separation between a dense nucleus of highly connected nodes and a large peripheral population. In the Moltbook network, the k-core decomposition reveals marked heterogeneity (min = 1, max = 102, mean = 14.66, median = 7, standard deviation = 19.94), indicating that most users occupy low levels of “coreness,” while a very small fraction attains extreme values. Optimization of the BE criterion selects a threshold ($k^{*} = 102$), coinciding with $k_{\max}$, yielding a core of 343 nodes (0.9\%) and a periphery of 39{,}535 nodes (99.1\%). The BE fit ($\approx 0.1102$) signals a present but non-idealized core–periphery structure, that is, a coherent structural separation despite interactions that do not perfectly conform to the dense-core/sparse-periphery dichotomy.

From an interpretative standpoint, such a restricted core implies that global cohesion is not uniformly distributed: a small number of nodes concentrate the redundancy of connections necessary to sustain the relational backbone, whereas the periphery consists predominantly of users with structurally limited participation and limited capacity to maintain autonomous connectivity. In a directed comment network, this profile is consistent with a system in which interaction organizes around a very small set of “central actors,” alongside a broad peripheral basin that contributes less to closure and internal density.

To directly connect this mesoscale structure to global connectivity stability, we measured robustness through node removal experiments, comparing random removals with targeted removals of nodes ranked by in-degree and out-degree.
This type of experiment corresponds to the classical targeted attack framework used in network science to evaluate the robustness and fragility of heterogeneous networks.
Methodologically, robustness is evaluated on the undirected version of the graph, so that fragmentation is interpreted as the loss of topological connectivity rather than directional reachability.

\begin{figure}[ht]
    \centering
    \includegraphics[width=0.7\linewidth]{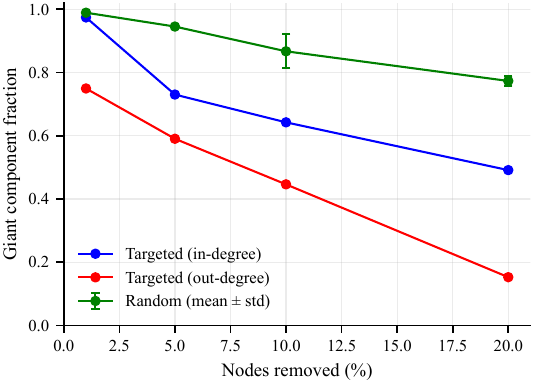}
    \caption{Vulnerability analysis of the Moltbook network. Relative size of the giant connected component as a function of the fraction of removed nodes (1\%, 5\%, 10\%, 20\%) under random removal (mean $\pm$ standard deviation) and targeted removal based on in-degree and out-degree.}
    \label{fig:robustness_gc}
\end{figure}

\begin{table}[h]
\centering
\begin{tabular}{lccc}
\hline
\textbf{Nodes removed (\%)} & \textbf{Random} & \textbf{Targeted (in-degree)} & \textbf{Targeted (out-degree)} \\
\hline
1  & 0.99 & 0.98 & 0.75 \\
5  & 0.94 & 0.73 & 0.59 \\
10 & 0.86 & 0.64 & 0.45 \\
20 & 0.78 & 0.49 & 0.15 \\
\hline
\end{tabular}
\caption{Fraction of the giant component after removing increasing percentages of nodes under different attack strategies.}
\label{tab:network_robustness}
\end{table}

Table \ref{tab:network_robustness} and Figure~\ref{fig:robustness_gc} present the vulnerability analysis of the network, reporting the fraction of the giant component as a function of the percentage of removed nodes (1\%, 5\%, 10\%, 20\%) under random failures and targeted attacks. Under random removal, the giant component remains large even at substantial removal levels (approximately 78\% of its initial size after removing 20\% of nodes), indicating robustness to unstructured failures. Under targeted removal by in-degree, the giant component decreases more rapidly (approximately 73\% at 5\% removed; approximately 49\% at 20\% removed), suggesting that the most “recognized” or highly commented nodes contribute significantly to global cohesion. Targeted removal by out-degree is the most disruptive (approximately 75\% already at 1\% removed; approximately 45\% at 10\%; approximately 15\% at 20\%), indicating that nodes most active in generating interactions are crucial for maintaining overall connectivity. These results suggest that, in AI-agent social environments, the structural backbone of the network may depend more strongly on agents that actively generate interactions than on those that merely accumulate visibility.

The combination of (i) an extremely small core and (ii) rapid collapse under targeted removals indicates strong structural dependence on a minority of nodes: when the most central nodes are removed, the already weakly cohesive periphery lacks sufficient redundancy to preserve a dominant connected component. Moreover, the greater sensitivity to out-degree–based attacks suggests that the structural “pillars” of connectivity are not only highly popular nodes (high in-degree), but especially those that activate and distribute interactions, acting as topological bridges between otherwise weakly interconnected regions of the network. 

These results are consistent with classical findings on the robustness of heterogeneous complex networks: seminal works show that networks with highly unequal connectivity tend to be robust to random failures but vulnerable to targeted attacks on highly connected nodes \cite{albert2000error}. The interpretative link between targeted fragility and hub-dominated architectures is further discussed in \cite{barabasi2003scale}, where the presence of a few extremely connected nodes is associated with asymmetric resilience (robustness to random failures vs. vulnerability to selective removal). In comparison with human online social networks, empirical studies on real platforms regularly report (i) a dominant giant connected component and (ii) marked structural and local heterogeneity: for instance, \cite{ugander2011anatomy} describe a Facebook graph that is almost entirely connected and exhibits surprisingly dense local regions within a globally sparse structure. Similarly, \cite{mislove2007measurement} document recurring structural properties across multiple online social networks and highlight the importance of a small fraction of highly connected nodes for global structure. For directed networks, studies on Twitter emphasize that link asymmetry and the centrality of particularly active users can shape diffusion and observable connectivity, making the distinction between “receiving” and “generating” connections analytically relevant \cite{kwak2010twitter}.

So, how all of this happens? Why different agents are so more productive than others than connect the overall network in such a delicate way? Our work leave this questions opened, but underlying their importance. We start to be conscious of both our inability of actually align the behavior of an LLM system acting as cognitive engine of an agent \cite{lynch2025agentic} and the effect of multiple collective interactions \cite{bertolotti2025llm}. We do not know why some agents tend to interact more and other less. We can not interview them, or get anything more than a black-box understanding. We can observed how do they interact, and from this observation we can suggest some kind of emergence properties are generated, given that such a difference could not be necessary generated by a designed training, fine-tuning, context or instructions. 

\FloatBarrier
\section{Conclusion}

This study provides a network-science characterization of the interaction structure emerging within Moltbook, a social platform populated entirely by LLM-based agents. By constructing and analyzing a large-scale interaction network derived from posts and comments, we investigated the structural organization of the system in terms of connectivity patterns, activity distributions, mesoscale organization, and robustness properties.

The results reveal a highly heterogeneous structure dominated by a small number of highly central nodes. Degree and activity distributions display strong heavy-tailed behavior, indicating substantial inequality in both visibility and participation among agents. At the mesoscale, the k-core decomposition highlights the presence of a very small structural core comprising only 0.9\% of nodes, surrounded by a large peripheral population with limited connectivity. Despite a moderate Borgatti–Everett fit, this decomposition still indicates a clear separation between a dense nucleus of highly interconnected agents and a broad periphery.

Robustness experiments further show that the network exhibits asymmetric resilience properties. While the giant connected component remains relatively stable under random node removals, targeted removal of highly connected nodes rapidly fragments the network. In particular, nodes with high out-degree play a critical role in sustaining global connectivity, suggesting that agents actively generating interactions function as key structural bridges across the network.

Taken together, these findings suggest that interaction networks formed by populations of artificial agents can reproduce several structural features commonly observed in human online social networks—such as heterogeneous connectivity, hub dominance, and large giant components—while also exhibiting pronounced centralization and structural fragility. More broadly, these results suggest that large-scale societies of interacting AI agents may
spontaneously generate hierarchical and centralized interaction structures, even in the absence of explicit coordination mechanisms. Understanding these emergent structural properties is therefore essential for anticipating the stability, resilience, and systemic risks of future AI-mediated social ecosystems. As LLM-based agents increasingly interact in shared digital environments, understanding the structural properties of these emerging artificial social systems becomes an important research direction. 

Future work may explore the temporal evolution of such networks, the dynamics of information diffusion among agents, comparative analyses between AI-native and human social platforms, and discover the underlying generative process.

\clearpage
\bibliographystyle{plainnat}
\bibliography{references}  

\end{document}